\documentclass[letterpaper,aastex]{emulateapj}
\tighten
\def\spose#1{\hbox to 0pt{#1\hss}}

\def\msun{{\rm ~M}_{\odot}}
\def\rsun{{\rm ~R}_{\odot}}

\def\mpy{{\rm ~M}_{\odot} {\rm ~yr}^{-1}}
\def\lta{\mathrel{\spose{\lower 3pt\hbox{$\mathchar"218$}}
     \raise 2.0pt\hbox{$\mathchar"13C$}}}
\def\gta{\mathrel{\spose{\lower 3pt\hbox{$\mathchar"218$}}
     \raise 2.0pt\hbox{$\mathchar"13E$}}}

\def\be{\begin{equation}}
\def\ee{\end{equation}}

\begin{document}

\title{Thermal Timescale Mass Transfer and the Evolution of 
White Dwarf Binaries}

\author{Natalia Ivanova \& Ronald E.~Taam}

\affil{  Northwestern University, Dept. of Physics \& Astronomy,
       2145 Sheridan Rd., Evanston, IL 60208\\
nata@northwestern.edu,  r-taam@northwestern.edu}

\begin{abstract}

The evolution of binaries consisting of evolved main sequence stars ($1 < M_d/\msun 
< 3.5$) with white dwarf companions ($0.7 < M_{wd}/\msun < 1.2$) is   
investigated through the thermal mass transfer phase. Taking into account the
stabilizing effect of a strong, optically thick wind from the accreting white dwarf
surface, we have explored the formation of several evolutionary groups of systems   
for progenitors with initial orbital periods of 1 and 2 days.  The numerical results 
show that CO white dwarfs can accrete sufficient mass to evolve to a Type Ia
supernova and ONeMg white dwarfs can be built up to undergo accretion induced
collapse for donors more massive than about $2 \msun$.  For donors less massive than
$\sim 2 \msun$ the system can evolve to form 
a He and CO or ONeMg white dwarf pair.  In addition,  
sufficient helium can be accumulated ($\sim 0.1 \msun$) in systems characterized 
by $1.6 \lta M_d/\msun \lta 1.9$ and $0.8  \lta M_{wd}/\msun \lta 1$ such that sub 
Chandrasekhar mass models for Type Ia supernovae, involving off center helium ignition,
are possible for progenitor systems evolving via the Case A mass transfer phase. 
For systems characterized by mass ratios $\gta 3$ the system likely merges as a result
of the occurrence of a delayed dynamical mass transfer instability.  A semi-analytical 
model is developed to delineate these phases which can be easily incorporated in 
population synthesis studies of these systems. 
\end{abstract}


\section{INTRODUCTION} 

Close binary systems consisting of main sequence-like stars with white dwarf 
companions have long been recognized as an important stage of evolution for understanding 
the formation of several diverse classes of objects.  Such systems are products of 
an evolution of a main sequence-like star with a red giant or asymptotic giant branch  
companion in which significant mass and orbital angular momentum have been lost.  The  
extremely non conservative evolution facilitated the transformation of the system from 
a wide orbit with an orbital period of $\sim 1$ year to a narrow one with a period 
of several days via a common envelope phase 
where the orbital energy released in the spiralling in process is sufficient to eject the 
common envelope (for reviews see Iben \& Livio 1993; Taam \& Sandquist 2000). 
For the orbital periods of post common envelope systems $\lta 0.5$ day, angular momentum 
losses by magnetic braking are effective (see Pylyser \& Savonije 1988) in shrinking 
the orbit further, producing cataclysmic variable systems.  On the other hand, for periods
$\gta 0.5$ day, the main sequence-like stars with masses $\gta 1 \msun$ can evolve to the 
mass transfer stage as a result of envelope expansion induced by the nuclear burning
in the main sequence star's interior.   The future evolution
of the nuclear evolved systems with donors in the mass range of $1 - 3 \msun$ 
may significantly contribute to the formation channels for the production of supersoft X-ray
sources (massive white dwarfs accreting at rates sufficient for steady hydrogen burning;
van den Heuvel et al. 1992), of a class of Type Ia supernovae models (Li \& van den Heuvel 
1997), of ultra short period (P $\lta 30$ min) interacting double white dwarf AM CVn 
systems (Podsiadlowski, Han, \& Rappaport 2003), and of detached double white dwarf systems 
(Nelemans et al. 2001). 

Common to the evolution of these systems is the occurrence of a phase of mass   
transfer on a thermal timescale.  Until recently, this population of systems was generally
neglected in population synthesis studies since it was assumed that the accreting white 
dwarf would expand to red giant dimensions 
as a result of reactivation of hydrogen burning at high mass accretion rates ($\gta 
10^{-7} \mpy$).  This expansion was hypothesized to lead to the formation of a second common 
envelope phase and to the eventual formation of a double degenerate system. However, it was pointed 
out by Kato \& Hachisu (1994) that an optically thick wind can be driven from the 
surface of white dwarfs more massive than about $0.5 \msun$, thereby stabilizing the 
mass transfer in the system and making new binary evolutionary channels possible.
In this case, the   
photosphere of the accreting white dwarf lies within its corresponding Roche lobe, 
and the system can be stabilized preventing evolution into the common envelope phase.
The existence of such solutions was made possible by a strong peak in the 
OPAL opacities at temperatures of about $1.6 \times 10^5$ K (Iglesias,
Rogers, \& Wilson 1987, 1990; Iglesias \& Rogers 1991, 1993; Rogers \& Iglesias 1992) 
achieved in the envelope of white dwarfs accreting at high rates. Recently, 
observational evidence for the accretion wind picture has been suggested by Hachisu 
\& Kato (2003) based on the long term variability of the light curve 
of the transient supersoft X-ray source RX J0513.9-6951 (see Reinsch et al. 2000).

Our study is, in part, similar to that conducted by Li \& van den Heuvel (1997), but 
differs in that we map out the boundaries 
delineating the evolutionary channels leading to the formation of double degenerate dwarfs, 
neutron stars by accretion induced collapse, and near Chandrasekhar and sub-Chandrasekhar Type 
Ia supernova models from such progenitor binary systems.  As such, we have carried out 
detailed binary evolutionary calculations of the Roche lobe filling donor at a stage 
of evolution between the main sequence and the base of the giant branch at orbital 
periods for which nuclear evolution rather than angular momentum losses dominate.
The evolution of the system is carried through the thermal timescale mass 
transfer phase to determine its ultimate outcome.  To obtain a clear picture of the 
possible evolutionary histories, the mass and evolutionary state 
of the donor star and the mass of the white dwarf are systematically varied for systems 
characterized by initial orbital periods of 1 and 2 days.   
The assumptions and input physics underlying our calculations 
are described in \S 2. The detailed binary sequences and numerical results are described
and compared with a semi-analytical picture for the boundaries of the various evolutionary 
channels in \S 3.  Finally, we summarize
and discuss the implications of our results in the concluding section. 
 
\section{FORMULATION}

The binary evolutionary sequences calculated in this investigation are based on 
a stellar evolution code developed by Kippenhahn, Weigert, \& Hofmeister (1967) 
and updated as described in Podsiadlowski, Rappaport, \& Pfahl (2002). The 
stellar models are computed using a reaction network with rates taken from 
Rauscher \& Thielemann (2000, 2001) (see also Thielemann, Truran, \& Arnould 1986) 
and using OPAL opacities (Rogers \& 
Iglesias 1992), supplemented with opacities at low temperatures (Alexander \&
Ferguson 1994), for a solar metallicity ($Z = 0.02$).  

\subsection{Mass transfer}

During phases of the evolution when the donor filled its Roche lobe, the mass 
transfer rate, $\dot M_{\rm tr}$, was calculated in an implicit manner. 
In this case, $\dot M_{\rm tr}$ is found such that
the radius of the donor is equal to its Roche lobe radius, $R_{\rm L}$
approximated as (see Eggleton 1983)
\begin{equation}
R_{\rm RL } = {\frac {0.49 q^{2/3} } {0.6 q^{2/3}+\ln (1+q^{1/3}) } } A,
\end{equation}
where $q$ is the ratio the mass of the donor, $M_d$, to the mass of the white dwarf, 
$M_{\rm wd}$, and $A$ is the orbital separation.  
We consider the radius-mass exponents of the Roche lobe
$\zeta_{\rm RL} = d \ln R_{\rm L}/d \ln M$
and of the star itself $\zeta= d \ln R/d \ln M$ in our solution method.
The response of the Roche lobe to the mass transfer (MT) is solely a function of
the mass ratio, whereas the response of the stellar radius to the MT
is the function of the MT rate.
For a given model, we tabulate values of $\zeta(\dot M)$.
By equating the Roche lobe radius and the predicted stellar radius,
one can find the MT rate for further models.
The MT rate solution is not necessarily unique since it can be multi-valued. 
In this case, we accept the smallest value of the MT rate.
While the star evolves, the response of the stellar radius evolves as well, and  
we recalculate the table of $\zeta(\dot M)$  if the predicted stellar
radius differs from the calculated one by 
$\delta \ln R = 10^{-4}$.

\subsection{Growth of the white dwarf}

The ultimate fate of the material transferred to the white dwarf depends on the 
mass and composition of the white dwarf and the rate of mass transfer.  
In this study, we consider CO white dwarfs with initial masses between $0.7 \msun$ 
and $1.15 \msun$ and ONeMg white dwarfs more massive than $1.15 \msun$.  The actual 
growth of the white dwarf is determined by the amount of accreted hydrogen 
which is eventually converted to helium and heavier elements.  In the case of 
rapid mass transfer, the white dwarf accumulates matter at the rate  
\begin{equation}
\dot M_{\rm cr} \sim 7.5 \times 10^{-7} (M_{\rm wd}/\msun - 0.4) \mpy,
\end{equation}
with the remaining matter radiatively driven away in a strong optically thick 
superwind from the white dwarf surface (Hachisu, Kato, \& Nomoto 1999).  The accumulation 
ratio, $\eta_H = \dot M_{\rm cr}/\dot M_{\rm tr} \leq 1$ . For 
transfer rates less than $\dot M_{\rm cr}$, but greater than about $\dot M_{\rm low} 
\sim 10^{-7} \mpy$, hydrogen burns steadily to helium ($\eta_H \sim 1$). 
If the mass transfer 
rate varies in the range between $\sim 3 \times 10^{-8} \mpy$ and $\dot M_{\rm low}$, 
the white dwarf experiences mild recurrent hydrogen shell flashes.
For the mass transfer rates below $\sim 3 \times 10^{-8} \mpy$,
strong unstable hydrogen shell flashes result in nova explosions, and we assume 
that the evolution is fully non conservative with all the transferred material 
ejected from the system.  In this regime, the white dwarf mass is constant (i.e., 
$\eta_H = 0$).  

In our calculations, $\eta_H$ is based on the work of Prialnik \& Kovetz (1995) who considered 
the accretion of hydrogen-rich matter onto white dwarfs of varying mass and 
central temperature. Since the accumulation ratios do not sensitively depend on the thermal 
state of the white dwarf, fits of their results were used  for accretion rates 
ranging from $10^{-9} M_\odot {\rm yr}^{-1}$ to $10^{-6} M_\odot {\rm yr}^{-1}$ 
for their hot white dwarf models.  

In the mass accretion rate regime where hydrogen-rich matter is retained, helium 
will naturally be accumulated.  
For a sufficiently massive helium layer the helium will be ignited.  
The critical mass necessary for helium ignition is dependent on 
the rate at which helium is processed in the hydrogen burning shell $\dot M _{\rm He}$ (Kato \& 
Hachisu 1999).  For rates less than $1.3 \times 10^{-6} \mpy$, helium shell burning 
is unstable in the white dwarf envelope. The accumulation ratio, defined as the ratio of the 
processed material remaining after one cycle of a helium flash to the ignition mass,  
has been estimated by Kato \& Hachisu (1999) as 
\begin{equation}
\eta_{\rm He} = \left\{ \begin{array} {l}
 1, \\
\ \ \ \ \ \ \ \ \ \ \ \ \ \ \ \ \ \ -5.9 \le \log \dot M _{\rm He}\lesssim -5 \\
-0.175(\log \dot M_{\rm He} + 5.35 )^2 +1.05), \\
\ \ \ \ \ \ \ \ \ \ \ \ \ \ \ \ \ \ -7.3 < \log \dot M_{\rm He} < -5.9
\end{array} \right.
\label{he_acc}
\end{equation} 
where $\dot M _{\rm He}$ is units of $M_\odot\ yr^{-1}$.
We note that the He shell is always unstable whenever a significant amount of hydrogen is 
processed to helium.  

If $\dot M _{\rm He}$ falls below the lower range in Eq.~\ref{he_acc}, the 
helium layer is not sufficiently hot to ignite helium shell burning in a thin mass layer.
In this case, helium is accumulated in the layer until the increasing density 
leads to the helium ignition in a thick mass layer of about 0.1 $\msun$. 
This ignition of the helium layer likely leads to the ignition of the CO or ONeMg core
disrupting the star as a sub-Chandrasekhar mass Type Ia supernova 
(see Taam 1980; Livne \& Glasner 1991; Woosley \& Weaver 1994; 
Garcia-Senz, Bravo, \& Woosley 1999).

\subsection{Orbital evolution}

The evolution of the orbital separation is dependent on the prescription for mass 
and angular momentum loss.  Since we consider systems characterized by orbital 
periods ($P \gta 1$ day) exceeding the bifurcation period (Pylyser \& Savonije 
1988) the only loss of orbital angular momentum from the system reflects that 
carried by the ejected matter.  In this study we assume that the mass lost in 
the radiatively driven wind carries the specific orbital angular momentum of 
the white dwarf.  Hence, the rate of angular momentum loss is given by 
\begin{equation}
\dot J = {M_d \over M_{wd}} {J \over M_d + M_{wd}} \dot M_{wind}
\end{equation}
where the orbital angular momentum, $J$, of the system is 
\begin{equation}
J = {M_d M_{wd} \over M_d + M_{wd}} A^2 {2 \pi \over P} 
\end{equation}
and $\dot M_{wind}$ is the mass loss rate from the system. 

The orbital and stellar evolution coupled to the stellar response to 
mass loss implicitly provides the constraint imposed on the system for determining 
the rate of mass transfer as described above.  

\section{RESULTS}

In order to facilitate an understanding of the detailed numerical results we, first,
approximately estimate the boundaries delineating the possible outcomes of the
evolution in the next subsection. A description of the detailed binary evolutionary 
sequences follows in the subsequent subsections. 

\subsection{Semi-analytical picture}

The evolutionary fate of a given initial binary system is critically dependent on the 
rate of mass transfer during the Roche lobe overflow phase.  As a result, 
an estimate for the average mass transfer rate is required to determine the approximate
boundaries separating the possible outcomes. In the following, we take the 
average mass transfer rate as
\begin{equation}
\overline {\dot M} = \Delta M  /t_{th},
\end{equation}
where $t_{th}$ is the thermal time-scale, on which the mass transfer operates, and
$\Delta M$ is the 
mass lost by the donor during this mass transfer event.
The thermal mass transfer from a MS star or a giant star takes place when the accretor is more 
massive star and terminates when 
$M_{\rm d }(t)=M_{\rm d }-\Delta M =M_{\rm wd}+ \eta_{\rm H} (\dot M)\eta_{\rm He}(\dot M_{\rm He}) 
\Delta M$.  In this description, only 
$\Delta M= (M_{\rm d }- M_{\rm wd })/(1+ \eta_{\rm H} 
(\dot M)\eta_{\rm He}(\dot M_{\rm He}) \eta_{\rm He})$ can be lost during the thermal mass transfer
phase.  In the case of the star evolving through the Hertzsprung Gap (HG), the thermal
time mass transfer is driven also by the star's own expansion,
and complete envelope can be lost ($\Delta M = M_{env}$) during this phase, however we do not treat it
separately in our simplified picture. 

The thermal timescale $t_{\rm th}$ on which $\Delta M$ is lost, is approximately given as
\begin{equation}
\begin{array} {l} 
t_{\rm th}
= t_{\rm th,\odot}  {\frac { M_{\rm d}/M_\odot \Delta M / M_\odot }
{R_{\rm d} / R_\odot \ L_{\rm d}/L_\odot}}  \\ 
\ \ \ \ = t_{\rm th,\odot}  {\frac { (M_{\rm d}/M_\odot)^{-2.25} \Delta M / M_\odot}
{R_{\rm d}/\rsun} }.\end{array}
\end{equation}
\noindent Here we used the mass-luminosity relation for main sequence stars, $L\propto M^{3.25}$, 
(this is also correct for
stars at HG, since they maintain the same luminosity as at the end of the MS), 
$ t_{\rm th,\odot}$ is the thermal time-scale of the Sun.
In this simplified estimate, we do not follow the evolution of the star through
the MT, and therefore use only the value of $t_{\rm th}$ at the onset of the MT.
The stellar radius, $R$, is identified then with the radius of the Roche lobe  
and is a function of the orbital period of the binary and the mass ratio of the system.
This yields a very simple description for the average mass transfer rate as

\begin{equation}
\overline {\dot M} = k{\frac {R_{\rm d}/\rsun(M_{\rm d}/M_\odot)^{2.25}  } {t_{\rm th,\odot}} },
\end{equation}

\noindent where k is some factor of the order of unity, which can be taken from the detailed 
calculations.

The final mass of the accretor is 
$M_{wd,f}=M_{wd} +\eta_{\rm H} (\dot M) \eta_{\rm He}(\dot M_{\rm He}) \overline {\dot M}t_{\rm th}$			
(for the complete description for $\eta_{\rm H}$ and  $\eta_{\rm He}$ see \S 2.2).  
This simple approximate approach permits an exploration of the parameter space for the 
final product of the binary evolution as a function of initial donor mass, white dwarf mass,  
and orbital period.  As representative examples we illustrate the boundaries for the possible 
evolutionary fates of systems characterized by initial orbital periods of 1 and 2 days in Figs. 
1 and 2 respectively.  For $P= 1$ day, where the thermal mass transfer occurs near the main 
sequence we adopt $t_{\rm th,\odot}\approx 3\times 10^7 $ yr.  On the other hand, 
a 1 $M_\odot$ star in the HG has a thermal timescale of $1.8\times 10^7 $ yr.  This corresponds 
to $t_{\rm th,\odot}\approx 6\times 10^7 $ yr in equation (7). 
For both Fig. 1 and 2 a value of $k=1.5$ was used based on the  
results of our detailed calculations

\begin{figure}
\plotone{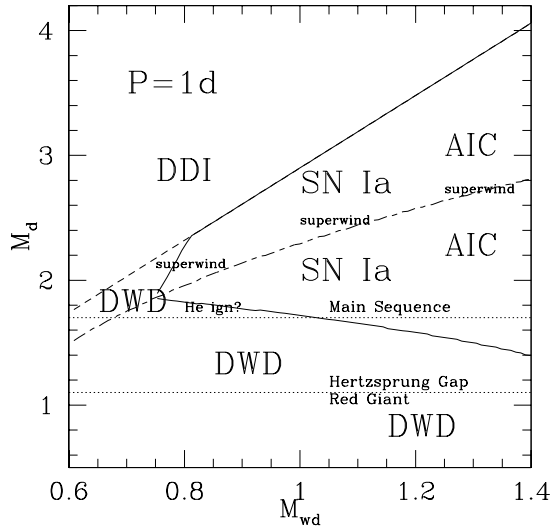}
\caption{Possible evolutionary fates of systems characterized by donors of initial 
mass $M_d$ and white dwarfs of initial mass $M_{wd}$ in units of $\msun$ for an orbital period of 
1 day based on a semi-analytical model (see text).  The solid curve delineates 
the area where the white dwarf can accrete sufficient mass for near Chandrasekhar mass 
models of SN Ia and AIC models of neutron star formation.  The dashed curve
represents the condition of the occurrence of a DDI (mass ratio q=3.1).  Double
white dwarf systems are formed at locations marked DWD. For reference, donors 
which undergo mass transfer above the upper dotted line lie close to the main sequence,
those below are in the Hertzsprung gap, and those below the lower dotted line are 
red giants.  Above the short dashed - long dashed line the mass transfer leads 
to the development of a superwind from the white dwarf. For stars that initiate 
mass transfer near the main sequence, but for which the rate is insufficient to 
lead to the superwind or to the SN Ia phase, a sub-Chandrasekhar SN is a possible outcome.}
\end{figure}
\begin{figure}
\plotone{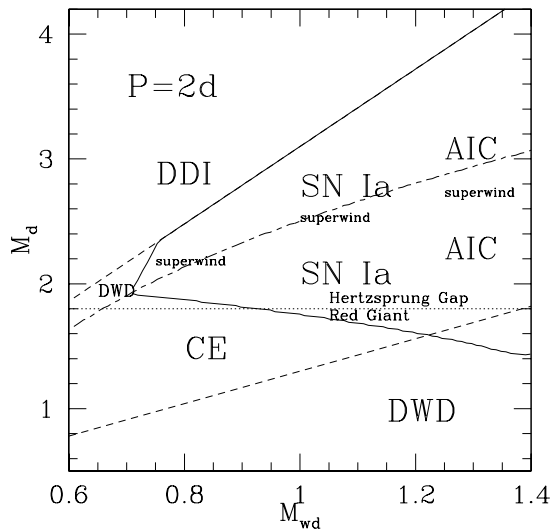}
\caption{Possible evolutionary fates of systems characterized by donors of initial 
mass $M_d$ and white dwarfs of initial mass $M_{wd}$ in units of $\msun$ for an orbital period of 
2 days based on a semi-analytical model (see text).  In addition to the boundaries 
described in Fig. 1, a lower dashed curve separates systems which enter into the 
common envelope phase (corresponding to mass ratios $\gta 1.3$) from those systems 
which evolve to form double white dwarf systems without requiring a common envelope phase.}
\end{figure}

The parameter range for which a white dwarf can accumulate sufficient mass to either 
evolve to a $1.4 \msun$ white dwarf required for a Type Ia supernova event (SN Ia) 
or an accretion induced collapse (AIC) 
is enclosed within the solid boundaries.  The initial mass of the white dwarf is 
the distinguishing characteristic in these evolutionary channels since CO white dwarfs
with initial masses $\lta 1.15 \msun$ explode upon the ignition of carbon, whereas ONeMg 
white dwarfs with initial masses $\gta 1.15 \msun$ collapse to form neutron stars as a 
result of electron capture processes.  The mass and composition of the initial 
white dwarf component of the system is necessary, but not sufficient to ensure 
these evolutionary fates, since the donor must supply matter at a rate
such that the white dwarf can grow significantly.  Only donors of an intermediate mass 
range can satisfy such a condition since the evolution is truncated for massive donors
by the onset of a delayed dynamical mass transfer instability (Webbink 1985; Hjellming 
1989) and for low masses by the occurrence of the nova phenomenon. Specifically, the upper 
limit to the donor's mass is a consequence of the change in the donor's response 
when the steep entropy profile of the outer envelope layers is removed.  In this 
case stellar contraction is replaced by stellar expansion as the flat entropy profile of 
the interior layers is exposed when sufficient matter is removed on a thermal timescale. Since the 
Roche lobe is contracting during this phase, the mass transfer process becomes dynamically
unstable.  Based on the results of our detailed calculations we adopt a critical mass ratio 
of 2.9 (for $P = 1$ day) and 3.1 (for $P = 2$ days) above which the delayed dynamical  
instability (DDI) is found.  For systems which enter the DDI
phase, it is highly likely that the system will merge, leaving behind a rapidly rotating  
remnant.  This upper limit, based on the DDI criterion, overestimates the donor's mass 
for systems characterized by white dwarfs
less massive than $1.05 \msun$ since donors in the mass range between $1.8 \msun$ and 
$3.3 \msun$ either are not sufficiently massive to build up the white dwarf to $1.4 \msun$ 
(even assuming 
an accumulation efficiency of unity) or because the efficiency of accretion is strongly reduced 
by the radiatively driven wind. A lower limit on the donor mass for the SN Ia and AIC 
evolutionary scenario is
determined by the occurrence of strong unstable hydrogen shell flashes in the envelope 
of the white dwarf.  In particular, the average mass transfer rate as determined by
the thermal timescale of the donor is insufficiently high for stars $\lta 1.7 \msun$
to prevent the ejection of significant mass from the system via nova explosions.

The remaining combinations of donor and white dwarf masses primarily lead to the  
formation of a double white dwarf (DWD) systems.
In these cases the mass transfer rate is 
insufficient to lead to the growth to the SN Ia or AIC phase.  We note that there exists
a narrow range of parameter space ($M_d \sim 1.7 \msun$ and $M_{wd} \sim 0.8 - 1 \msun$) 
for which a significant layer of helium may be accumulated, giving rise to an 
evolutionary channel in which a sub-Chandrasekhar mass supernova model may be viable.

The range in donor and white dwarf masses delineating the evolutionary fates of systems 
in Fig. 2 for an initial orbital period of 2 days at the onset of mass transfer are
similar to those described in Fig. 1, allowing for the possibility that white dwarfs can
be significantly built up.  For this greater orbital period, the main sequence 
like donors are evolved to a greater extent and the onset of mass transfer occurs after 
the formation of a helium core when the donor is in the Hertzsprung gap or at the 
base of the giant branch. As is evident 
from Fig. 2 there exists a region where the system can enter into a 
common envelope phase as a result of evolution to the giant stage. The lower 
dashed line denotes this region corresponding to the condition that the 
mass ratio equals 1.3 as based on our detailed numerical calculations.  For 
mass ratios greater than 1.3 the system enters into 
the common envelope phase.  In this case the donor has a well defined core-envelope 
structure typical of red giant stars.  A double degenerate system consisting of 
a He white dwarf with a CO or ONeMg white dwarf in a short period orbit ($P \lta 1$  
hour) may result depending on the particular evolutionary state of the donor
at onset of the common envelope stage (Sandquist, Taam, \& Burkert 2000).  
An additional difference between the systems depicted in Fig. 1 and 
in Fig. 2 is the absence of systems which can evolve to a sub-Chandrasekhar 
mass SN.  This reflects the fact that the regime where a sufficient helium
mass layer accumulates on the white dwarf occurs for systems in which the mass 
transfer takes place 
only when the donor is close to the main sequence.  For donors
of more advanced evolutionary stages in mass transferring systems at initial
orbital periods of 2 days, the timescale for evolution is so short that significant 
accumulation of helium required for a sub Chandrasekhar model is not found.

\subsection{Detailed calculations}

We have computed the evolution of binary systems initially consisting of main 
sequence-like stars of masses 1 - $3.8 \msun$ and white dwarfs of masses $0.7 - 1.2 
\msun$ through the thermal mass transfer phase.  The evolutionary state of the
donor stars was chosen such that the Roche lobe overflow phase was initiated at
either an orbital period of 1 or 2 days.  To sample the parameter space adequately,
a total of 65 evolutionary sequences were calculated with 28 and 37 sequences 
calculated for systems at initial orbital periods of 1 and 2 days respectively.
The systems were evolved until the accreting white dwarf had reached $1.4 \msun$, 
the donor had evolved to a point where it 
was clear that the system would evolve into a DWD system, 
or the system enters into a common envelope phase as a 
result of either a dynamical or delayed dynamical mass transfer instability.

\begin{figure}
\plotone{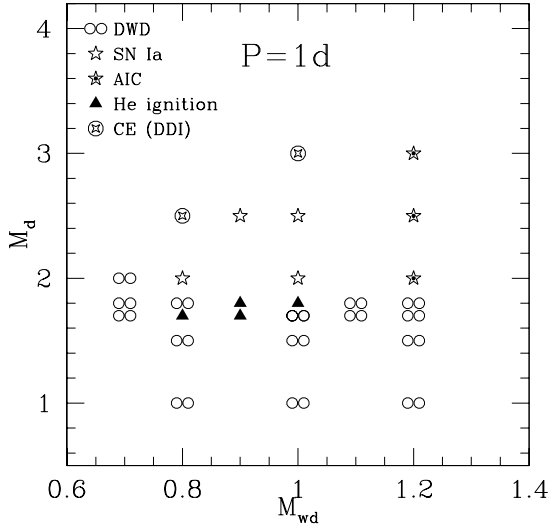}
\caption{The evolutionary fates of the binary sequences calculated for an initial 
orbital period of 1 day 
and donors of initial 
mass $M_d$ and white dwarfs of initial mass $M_{wd}$ in units of $\msun$. 
The symbols representing the fate of the system as a double 
white dwarf (DWD), a near Chandrasekhar mass supernova model (SN Ia), a helium ignition sub
Chandrasekhar model (He ignition), accretion induced collapse (AIC), and an evolution 
into the common envelope stage 
via a delayed dynamical mass transfer instability (CE DDI) are shown.}
\end{figure}
\begin{figure}
\plotone{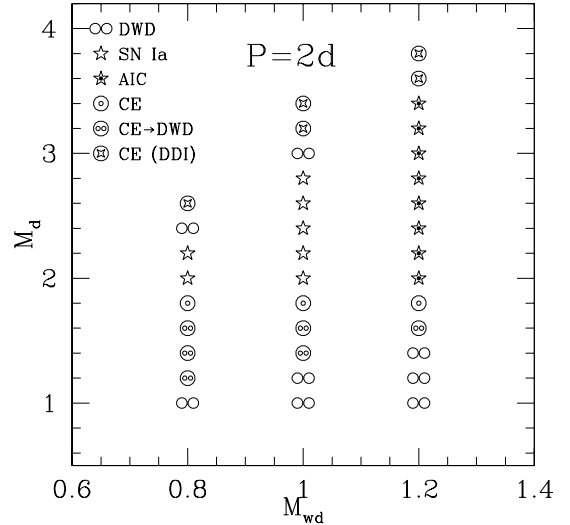}
\caption{The evolutionary fates of the binary sequences calculated for an initial 
orbital period of 2 days 
and donors of initial 
mass $M_d$ and white dwarfs of initial mass $M_{wd}$ in units of $\msun$. 
The symbols representing the fate of the system as a double 
white dwarf (DWD), a near Chandrasekhar mass supernova model (SN Ia), accretion induced 
collapse (AIC), and an evolution into the common envelope stage via a dynamical instability
(CE) leading to merger and to possibly a DWD system (CE $\rightarrow$ DWD), and delayed dynamical 
mass transfer instability (CE DDI) are shown.}
\end{figure}

A visual summary of the fate of all the calculated evolutionary sequences is 
presented in Figs. 3 and 4 for initial orbital periods of 1 and 2 days 
respectively.  Upon inspection of the boundaries displayed in Figs. 1 - 4, 
it is clear that the results based on the semi-analytical approach compare  
very favorably with the detailed binary evolutionary computations. 
The quantitative  results of representative evolutionary sequences are listed in Table 1
and 2 for initial orbital periods of 1 and 2 days respectively. 
Here, the initial mass of the white dwarf, $M_{wd;0}$, mass of the white dwarf at 
the end of the calculations,  
$M_{wd;f}$, 
the initial mass of the donor, $M_{d;0}$, mass of the donor at the end of the 
calculations, $M_{d;f}$, 
the minimum orbital period that the binary system has reached during the computed evolution,
$P_{\rm min}$,
the  orbital period at the end of the calculations, $P_f$, 
the evolution time from the onset of the mass transfer  
phase, $t_{tr}$, and
the average mass transfer rate, $\dot M_{tr}$ are listed. 

In the following, we divide the description of the detailed numerical results into 
those sequences for which growth of the accreting white dwarf is sufficient to lead to 
a near Chandrasekhar or sub-Chandrasekhar mass model for SN Ia and to an AIC phase and
sequences leading to the formation of double degenerate systems or to a merged object.

\subsubsection{Growth of white dwarf to SN Ia or AIC phase}

The progenitor systems which favor the growth of the white dwarf to a 
near Chandrasekhar mass are characterized by massive white dwarfs ($M_{\rm wd;0}
\gta 0.8 \msun$).  It can be inferred from Tables 1 and 2 that the efficiency 
of mass accretion by the white dwarf, as defined by the ratio of the 
mass accreted by the white dwarf to the mass lost by the donor, is a strong 
function of the mass and evolutionary state of the donor.  Specifically,
the efficiency of white dwarf growth ranges from about 10\% to 70\% over the
calculated grid with the highest efficiencies obtained for donors with  
masses $\sim 2 \msun$.  Clearly the mass transfer rates for these donors 
are in the regime where steady hydrogen burning takes place, but yet the 
superwind is not so effective in driving a significant amount of matter from 
the system.  That is, the thermal timescale mass transfer associated with 
donors of $\sim 2 \msun$ favors a higher efficiency of material accumulation on 
the white dwarf in comparison to other donors.
In contrast, the progenitor systems which lose a significant 
fraction of their mass via the radiatively driven wind from the white 
dwarf surface during the evolution generally are characterized by 
systems with massive donors $\gta 2.5 \msun$.  In fact, the degree to 
which systems lose mass is exemplified by the systems which evolve to the 
SNIa or AIC phase.  For these systems, the total mass loss from the  
system can lead to the presence of $\sim 0.4 - 1.9 \msun$ surrounding the 
system at the time the white dwarf has been built up to $1.4 \msun$.

\begin{figure}
\plotone{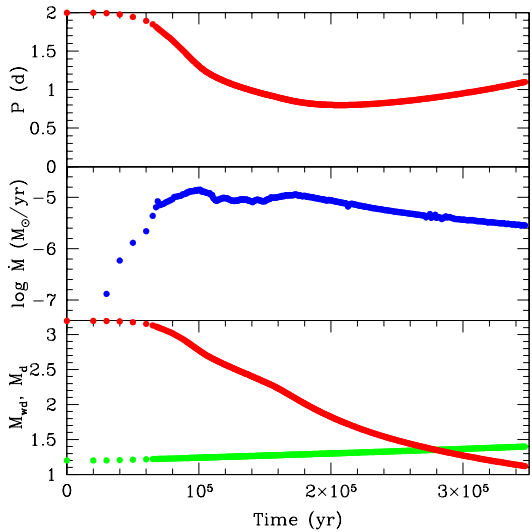}
\caption{Evolution of a binary characterized by $M_{d;0} = 3.2 \msun$, 
$M_{wd;0} = 1.2 \msun$ at an initial orbital period of 2 days. Upper panel: 
The temporal evolution of the orbital period. Middle panel: The temporal 
evolution of the mass transfer rate. Lower panel: The variation of the mass of the donor
and white dwarf during the evolution.}
\end{figure}

An example of such an evolution characterized by an ONeMg white dwarf of 
$M_{wd;0} = 1.2 \msun$ and a donor of 
$M_{d;0} = 3.2 \msun$ at an initial orbital period of 2 days is 
illustrated in Fig. 5. In this case the mass transfer rates rise 
rapidly to $\sim 10^{-5} \mpy$ within $10^5$ yrs and averages about $6.8 
\times 10^{-6} \mpy$ over the entire evolution.  
The efficiency of the mass accretion onto 
the white dwarf is low in this sequence amounting to $\sim 10\%$.
The orbital period initially decreases by more than a factor of 2 
reaching a minimum period of $0.8$ days after $2 \times 10^5$ yrs 
before increasing to a period of 1.1 day at which point the white 
dwarf has increased to $1.4 \msun$ and the donor has decreased to 
$1.1 \msun$.  The mass loss in such a sequence has been extensive 
with about $1.9 \msun$ enveloping the system. It is expected that 
the system would undergo an accretion induced collapse further 
widening the orbit as the gravitational mass of the compact object 
is reduced by about $0.1 - 0.2 \msun$.  With the additional 
expansion of the donor, the system will evolve to longer orbital 
periods and enter the low mass X-ray binary phase. 

We have also found that systems with donors and white dwarfs in a 
narrow range ($1.6 \lta M_{d;0}/ \msun \lta 1.9$, $0.8 \lta M_{wd;0}/ 
\msun \lta 1$) can lead to the accumulation of a sufficient layer of 
helium mass on the CO white dwarf required for the initiation of an off center 
helium detonation.  The conditions required for the growth of the 
helium layer are a function of the evolutionary state of the donor 
since the strong helium flash regime was found to only occur for donors 
which undergo the MT phase near the main sequence at short orbital
periods ($P \sim 1$ day). 
 
\subsubsection{Formation of DWD or single merged object}

Although the evolutionary sequences were carried out only through the thermal  
timescale mass transfer phase, the outcome of an evolution as a DWD system can be inferred  
once the mass transfer rates in the system have decreased to the point when  
strong hydrogen shell flashes are expected to occur in the white dwarf envelope.  
Figures 3 and 4 illustrate that DWD systems can form from progenitor systems with either 
a low mass ($\lta 1.4 \msun$) or high mass ($\gta 2.6 \msun$) donor. For the lower
mass donors, the temporal evolution of the mass transfer rates lead to nova explosions
over part of the evolution such that the white dwarf does not build up to $1.4 \msun$.
On the other hand, there is a narrow range for high mass donors where, although the white dwarf
can be built up significantly, the donor becomes less massive than the white 
dwarf during the evolution such that the mass transfer rates have decreased to the 
extent that accretion is 
prevented by the occurrence of nova explosions. The DWD systems formed via this channel
are expected to have long orbital periods ($P \gta 1$ days).
 
The evolutionary channel involving a CE phase may also produce DWD systems.  The 
calculations reveal that this evolutionary channel is restricted to the mass 
transfer taking place after the donor has evolved through the Hertzsprung 
gap.  The response of the star with a deep convective envelope leads to a 
mass transfer instability and to the evolution into the CE phase. The outcome
of this evolution is not well understood, but if the system survives it is likely 
to have a short orbital period ($P \lta 1$ hr). The common envelope calculations 
carried out by Sandquist, Taam, \& Burkert (2000) indicate that evolution into
the CE near the base of the giant branch may lead to the successful 
ejection of the common envelope if the degenerate helium core is more massive than 
$\sim 0.2 - 0.25 \msun$. 

\begin{figure}
\plotone{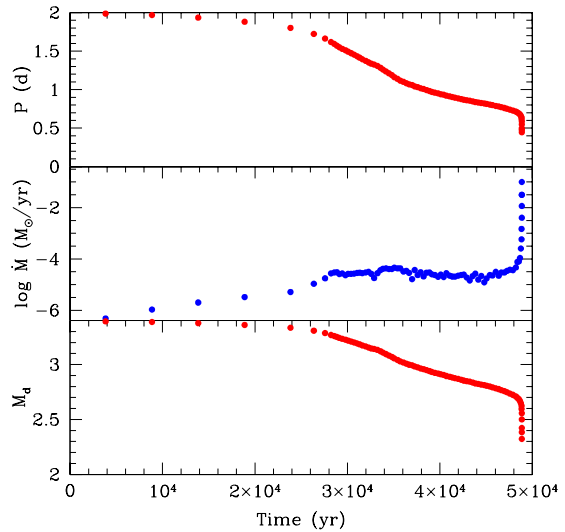}
\caption{Evolution of a binary characterized by $M_{d;0} = 3.4 \msun$, 
$M_{wd;0} = 1 \msun$ at an initial orbital period of 2 days for a system that 
evolves to a delayed dynamical mass transfer instability. Upper panel: 
The temporal evolution of the orbital period. Middle panel: The temporal 
evolution of the mass transfer rate. Lower panel: The variation of the mass of the donor
during the evolution. The white dwarf mass is not shown since it increases 
by only $0.02 \msun$.}
\end{figure}

For more massive donors, the thermal timescale mass transfer phase enters into a period 
where the mass transfer rate accelerates rapidly, 
indicating the onset of the delayed dynamical mass transfer instability.  For 
example, a sequence characterized by $M_d = 3.4 \msun$, $M_{wd} = 1 \msun$ at 
an initial orbital period of 2 days is illustrated in Fig. 6. The mass transfer rate accelerates by 
three orders of magnitude after $\sim 4.9 \times 10^4$ years with the rapid 
increase occurring over a timescale of $\lta 200$ years. 
In these sequences, the mass transfer initially takes place on a thermal timescale 
of the donor until the outer radiative envelope layer has been removed. Further 
mass loss leads to the exposure of the layers characterized by a flat entropy 
profile and expansion of the donor results. 
Since the Roche lobe during this phase is contracting, the system evolves
into a CE phase.  In this case, merger of the system is likely.  The occurrence of 
this instability occurs at a mass ratio of about 3.1 for donors in the Hertzsprung 
gap and about 2.9 for donors near the main sequence, confirming earlier estimates  
by Hjellming (1989).

\section{SUMMARY}

The fate of binary systems which undergo a phase of mass transfer on a thermal    
timescale has been investigated for binaries composed of main sequence-like donors with
white dwarf companions.  Allowing for the possibility of an optically thick wind 
driven from the white dwarf surface gives rise to evolutionary channels in which the 
white dwarf accretes sufficient mass as required for either a 
near Chandrasekhar mass or sub-Chandrasekhar mass model for Type Ia supernovae or 
the formation of neutron stars via an
accretion induced collapse model.  The range for donor and white dwarf masses  
delineating the formation of these objects from those producing 
a pair of degenerate dwarfs has been determined by detailed computations, and 
reproduced by a simple semi-analytic model, for progenitor systems characterized
by initial orbital periods of 1 and 2 days.  

Specifically, we have found that CO white dwarfs can accumulate sufficient 
matter to produce conditions ripe for initiation of a central carbon deflagration/detonation 
supernova explosion 
provided that their initial masses are greater than about $0.8 \msun$ and their donor is 
more massive than about $2 \msun$.  An upper limit to the donor mass is set by the 
onset of a delayed dynamical instability which occurs at a mass ratio of $\sim 3$. 
In addition, the ONeMg white dwarfs (with 
masses $\gta 1.15 \msun$) may accrete sufficient matter to collapse to form  
a neutron star by electron capture processes for binary systems with donors in the 
same mass range.  During these phases, the mass transfer rates are sufficiently 
high that hydrogen burning provides the bulk of the energy generation such that 
the sources are likely to be observed as supersoft X-ray sources (van den Heuvel 
et al. 1992; Kahabka \& van den Heuvel 1997). These evolutionary  
results are similar to those reported by Li \& van den Heuvel (1997)
and extend the lower bound of the progenitor white dwarf masses from $0.9 \msun$ to 
about $0.8 \msun$ and less, confirming similar results obtained by Langer et al. (2000)
\footnote{In a very recent study on SN~Ia progenitors an even lower bound for WD
masses was reported -- 0.67 $M_\odot$ (Han \& Podsiadlowski,
astro-ph/0309618). This value, however, was based upon the full accretion
of hydrogen rich material even for $\dot M < \dot M_{\rm low}$.}.  
Although Li \& van den Heuvel (1997) did not consider accretion induced collapse
in their study, their results are also applicable for this neutron star formation 
path as well.  In these cases, the amount of matter surrounding the system as 
a result of the wind loss from the white dwarf surface may be as high as $1 - 2 \msun$.
The existence of such circumbinary material is not dissimilar to that expected for 
complementary evolutionary channels leading to a SN Ia involving mass transfer from 
an asymptotic giant branch star to its massive white dwarf companion. We note that 
additional  hydrogen rich matter may also be lost from the system resulting from the interaction
of the supernova shell with the donor (Marietta, Burrows, \& Fryxell 2000).  
In the past, one of 
the arguments put forth against such accreting models is the lack of hydrogen emission
in the spectra of SN Ia.  However, the recent discovery of hydrogen emission in  
the observations of the type Ia supernova SN2002ic 
by Hamuy et al. (2003) provides some support for such models.  Although 
Hamuy et al. (2003) attribute the hydrogen emission from circumstellar matter
surrounding an asymptotic giant branch star, the wind present in the accreting
white dwarf model in the short period systems investigated here suggests that 
sufficient matter surrounding the system may be a characteristic of these models as well.

Our detailed systematic investigation has also led to an identification of a 
channel where a sufficient helium-rich mass layer accumulates ($\sim 0.1 
\msun$) below the hydrogen-rich layer such that an off center helium detonation 
may be ignited in the white dwarf envelope.  The propagation of the nuclear burning 
front into the core region, enhanced by geometrical focusing, may lead to the 
incineration of the entire white dwarf (Woosley \& Weaver 1994) for a sub-Chandrasekhar 
mass star.  The range of masses of the progenitor systems as well as the evolutionary 
state of the donor which follow this evolutionary channel are narrow in the parameter 
range studied in this paper.  That is,  
only donors in the mass range between 1.6 and  1.9 $\msun$ which transfer matter to 
their white dwarf companion (with masses in the range $0.8  \lta M_{wd}/\msun \lta 1$) 
while close to the main sequence are viable.  Such limited properties of the progenitor
systems may be consistent with the fact that such models are expected to be minor 
contributors to the SN Ia rate since these models are subluminous compared to 
models based on the carbon deflagration/detonation near Chandrasekhar mass models. 

For donors less massive than $\sim 2 \msun$ less matter is transferred  
to the white dwarf, leading to systems composed of a He white 
dwarf with a CO or ONeMg white dwarf companion. Of these systems, those that 
undergo a dynamical mass transfer instability evolve into the common envelope  
phase.  For those systems which survive, the systems are likely to emerge at  
short orbital periods ($P \lta 1$ h), providing an alternative evolutionary 
channel for the formation of AM CVn binary systems in addition to the channel involving the
stable mass transfer evolution of evolved secondaries in cataclysmic variable  systems 
(see Podsiadlowski, Han, \& Rappaport 2003).  On the other hand, the
systems which do not enter into the 
common envelope phase produce double degenerate systems characterized by orbital 
periods which are greater than about 1 day. 

The systems which undergo an accretion induced collapse likely evolve  
to a low mass X-ray binary phase in which a main sequence-like star with an evolved 
core transfers mass to its newly formed neutron star companion. Investigations
of this phase have recently been carried out by Sutantyo \& Li (2000), who considered  
the accretion induced collapse scenario, as well as the intermediate mass X-ray 
binary scenario (also studied by Podsiadlowski et al. 2002).  The system 
evolves to become a binary millisecond pulsar with a helium white dwarf companion 
in a short period system. Our results suggest that such systems which form via AIC 
process are likely to have orbital periods greater than about 1 day. 

Finally, we point out that the semi-analytical model for the white dwarf binaries 
that we have developed in this paper can be easily incorporated into population 
synthesis investigations.  We plan to carry out such calculations in the future 
in order to assess  the importance of the formation 
channels associated with the thermal timescale mass transfer phase in these 
systems. 

\acknowledgments
We thank the referee for comments which improved the clarity of this paper.  This work is 
partially supported by the NSF through grant AST-0200876.

\vfil\eject

\begin{deluxetable}{l l l l l l l l l}
\tabletypesize{\scriptsize}
\tablecaption{
The binary parameters and outcomes of representative model sequences for 
systems in which the onset of mass transfer occurs at a orbital period of
1 day. }
\tablecomments{The columns denote: the mass of the white dwarf, $M_{\rm wd;0}$ and of
the donor $M_{\rm d; 0}$ at the onset of the MT (in $\msun$); the mass 
of the white dwarf $M_{\rm wd;f}$ (in $\msun$), the mass of the donor $M_{\rm d; f}$
(in $\msun$), and the period $P_{\rm f}$ (in days) at the end of the
computation; $P_{\rm min}$ (in days) is the minimum period of the binary
system during the computed interval of evolution, $\Delta t_{\rm tr}\ [yr] $
is the time of the MT phase, and $\dot M_{\rm tr}$ (in $\mpy$) is the corresponding
average MT rate. The possible 
outcomes are: DWD: double white dwarf system,
SNIa - supernova type Ia, subCh - sub-Chandraskehar supernova,
AIC - accretion-induced collapse.}
\tablehead{
\colhead{$M_{\rm wd;0}$}    & \colhead{$M_{\rm d; 0}$} &
\colhead{$M_{\rm wd; f}$}   & \colhead{$M_{\rm d; f}$}  &
\colhead{$P_{\rm min}\ [d]$} & \colhead{$P_{\rm f}\ [d]$}   & \colhead{$\Delta t_{\rm tr}\ [yr] $} &
\colhead{$\dot M_{\rm tr}\ [M_\odot /yr]$}   &\colhead{outcome}   }
\startdata
0.7  &  2.0   &  1.34   &  0.89  &  0.42 & 0.5   &  5.5$\cdot 10^6$  & $2\cdot
10^{-7}$ & DWD  \\
0.8  &  1.5   &  0.81   &  0.75  &  0.82 &  1.02  &  7.9$\cdot 10^7$  & $9.5\cdot
10^{-9}$ & DWD  \\
0.8  &  1.7   &  0.9    &  1.25  &  0.69 &  0.69  &  9.5$\cdot 10^6$  & $4.7\cdot
10^{-8}$ & subCh\\
0.8  &  2.0   &  1.4    &  1.17  &  0.52 &  0.55  &  3.9$\cdot 10^6$  & $2.2\cdot
10^{-7}$ & SNIa \\
1.0  &  1.0   &  1.02   &  0.90  &  1.0 &  1.02  &  1.1$\cdot 10^7$  & $9\cdot
10^{-9}$ & DWD  \\
1.0  &  2.5   &  1.4    &  1.37  &  0.51 &  0.51  &  1.7$\cdot 10^6$  & $6.6\cdot
10^{-7}$ & SNIa \\
1.2  &  2.0   &  1.4    &  1.70  &  0.86 &  0.86  &  3.3$\cdot 10^6$  & $9.1\cdot
10^{-8}$ & AIC  \\
1.2  &  3.0   &  1.4    &  1.64  &  0.5 &  0.5   &  2.9$\cdot 10^6$  & $4.7\cdot
10^{-7}$ & AIC  \\
\enddata
\label{table}
\end{deluxetable}

\begin{deluxetable}{l l l l l l l l l}
\tabletypesize{\scriptsize}
\tablecaption{The binary parameters and outcomes of representative model
sequences for systems in which the onset of mass transfer occurs at  
an orbital period of 2 days. }
\tablecomments{ The columns are the same as for Table 1;
the additional possible outcome, 
CE$ \rightarrow $DWD, corresponds to systems that possibly survive the common 
envelope phase and form
a short period double white dwarf system.}
\tablehead{
\colhead{$M_{\rm wd;0}$}    & \colhead{$M_{\rm d; 0}$} &
\colhead{$M_{\rm wd; f}$}   & \colhead{$M_{\rm d; f}$}  &
\colhead{$P_{\rm min}\ [d]$} &\colhead{$P_{\rm f}\ [d]$}   & \colhead{$\Delta t_{\rm tr}\ [yr] $} &
\colhead{$\dot M_{\rm tr}\ [M_\odot /yr]$}   &\colhead{outcome}   }
\startdata
0.8  &  2.0   &  1.4    &  1.25  &  1.05 &  1.08   &  2.$\cdot 10^6$  & $3.8\cdot
10^{-7}$ & SNIa  \\
0.8  &  2.4   &  1.32   &  0.32  &  0.58 &  4.2   &  9.1$\cdot 10^6$  & $2.3\cdot
10^{-7}$ & DWD   \\
1.0  &  1.4   &  1.0    &  0.2   &  2.0 &  0.02\tablenotemark{a}  &                   &
& CE$ \rightarrow $DWD \\
1.0  &  2.0   &  1.4    &  1.46  &  1.44&  1.44   &  2.6$\cdot 10^6$  & $2.1\cdot
10^{-7}$ & SNIa  \\
1.0  &  2.4   &  1.4    &  1.4   &  1.1 &  1.1   &  8.4$\cdot 10^5$  & $1.2\cdot
10^{-6}$ & SNIa  \\
1.0  &  2.8   &  1.4    &  0.6   &  0.7 &  1.7   &  7.7$\cdot 10^5$  & $2.9\cdot
10^{-6}$ & SNIa  \\
1.2  &  2.0   &  1.4    &  1.67  &  1.7 &  1.7   &  2.5$\cdot 10^6$  & $1.3\cdot
10^{-7}$ & AIC   \\
1.2  &  2.6   &  1.4    &  1.97  &  1.4 &  1.4   &  3.6$\cdot 10^5$  & $1.7\cdot
10^{-6}$ & AIC   \\
1.2  &  3.2   &  1.4    &  1.1   &  0.8 &  1.1   &  3.1$\cdot 10^5$  & $6.8\cdot
10^{-6}$ & AIC   \\
\enddata
\label{numbers}
\tablenotetext{a} {determined by equating the orbital energy to the 
binding energy of the donor's envelope}
\end{deluxetable}


\begin{thebibliography}{}

\bibitem [a (1900)]{a}Alexander, D. R., \& Ferguson, J. W. 1994, \apj, 437, 879
\bibitem [b (1900)]{b}Eggleton, P. P. 1983, \apj, 268, 368
\bibitem [c (1900)]{c}Garcia-Senz, D., Bravo, E., \& Woosley, S. E. 1999, \aap, 349, 177
\bibitem [d (1900)]{d}Hachisu, I., Kato, M., \& Nomoto, K. 1999, \apj, 522, 487
\bibitem [e (1900)]{e}Hachisu, I., \& Kato, M. 2003, \apj, 590, 445
\bibitem [xx (1900)]{f}Hamuy, M. et al. 2003, \nat, in press
\bibitem [xx (1900)]{g}Hjellming, M.S. 1989, Ph.D. thesis, University of Illinois
\bibitem [xx (1900)]{i}Iben, I. Jr., \& Livio, M. 1993, \pasp, 105, 1373
\bibitem [xx (1900)]{j}Iglesias, C. A., \& Rogers, F. 1991, \apj, 371, L73
\bibitem [xx (1900)]{k}Iglesias, C. A., \& Rogers, F. 1993, \apj, 412, 752
\bibitem [xx (1900)]{l}Iglesias, C. A., Rogers, F. J., \& Wilson, B. G. 1987, \apj, 322, L45
\bibitem [xx (1900)]{m}Iglesias, C. A., Rogers, F. J., \& Wilson, B. G. 1990, \apj, 360, 221
\bibitem [xx (1900)]{n}Kahabka, P., \& van den Heuvel, E. P. J. 1997, \araa, 35, 69
\bibitem [xx (1900)]{o}Kato, M., \& Hachisu, I. 1994, \apj, 437, 832
\bibitem [xx (1900)]{p}Kato, M., \& Hachisu, I. 1999, \apj, 513, L41
\bibitem [xx (1900)]{r}Kippenhahn, R., Weigert, A., \& Hofmeister, E. 1967,
in Methods in Computational Physics, Vol. 7,
ed. B. Alder, S. Fernbach, \& M. Rothenberg (New York: Academic), 129
\bibitem [xx (1900)]{s}Langer, N., Deutschmann, A., Wellstein, S., \& H\"oflich, P. 2000, 
\aap, 362, 1046
\bibitem [xx (1900)]{t}Li, X. D., \& van den Heuvel, E. P. J. 1997, \aap, 322, L9
\bibitem [xx (1900)]{u}Livne, E., \& Glasner, A. S. 1991, \apj, 370, 272
\bibitem [xx (1900)]{x}Marietta, E., Burrows, A., \& Fryxell, B. 2000, \apjs, 128, 615
\bibitem [xx (1900)]{y}Nelemans, G. Yungelson, L. R., Portegies Zwart, S. F., \& Verbunt, 
F. 2991, \aap, 365, 491
\bibitem [xx (1900)]{z}Podsiadlowski, Ph., Han, Z., \& Rappaport, S. 2003, \mnras, 340, 1214
\bibitem [xx (1900)]{qq}Podsiadlowski, Ph., Rappaport, S., \& Pfahl, E. 2002, \apj, 565, 1107
\bibitem [xx (1900)]{aa}Prialnik, D., \& Kovetz, A. 1995, \apj, 445, 789
\bibitem [xx (1900)]{zz}Pylyser, E.~\& Savonije, G.~J.\ 1988, \aap, 191, 57
\bibitem [xx (1900)]{xx}Rauscher, T., \& Thielemann, F.-K. 2000, 
Atomic Data and Nuclear Data Tables, 75, 1
\bibitem [xx (1900)]{ss}Rauscher, T., \& Thielemann, F.-K. 2001, 
Atomic Data and Nuclear Data Tables, 79, 47
\bibitem [xx (1900)]{ww}Reinsch, K., van Teeseling, A., King, A. R., \& Beuermann, K. 2000, 
\aap, 354, L37
\bibitem [xx (1900)]{ee}Rogers, F. J., \& Iglesias, C. A. 1992, \apjs, 79, 507
\bibitem [xx (1900)]{dd}Sandquist, E. L., Taam, R. E., \& Burkert, A. 2000, \apj, 533, 984
\bibitem [xx (1900)]{cc}Sutantyo, W. \, \& Li, X. D. 2000, \aap, 360, 633
\bibitem [xx (1900)]{vv}Taam, R. E. 1980, \apj, 242, 749
\bibitem [xx (1900)]{ff}Taam, R. E., \& Sandquist, E. L. 2000, \araa, 38, 113
\bibitem [xx (1900)]{rr}Thielemann, F.-K., Truran, J. W., \& Arnould, M. 1986,
in Advances in nuclear astrophysics, ed. Frontieres (Gif-sur-Yvette, France), 525
\bibitem [xx (1900)]{tt}van den Heuvel, E. P. J., Bhattacharya, D., Nomoto, K., \& Rappaport, 
S. A. 1992, \aap, 262, 97
\bibitem [xx (1900)]{gg}Webbink, R. 1985, in Interacting Binary Stars, ed. J.E.~Pringle \& R.A.
Wade, (Cambridge:Cambridge University Press), 39
\bibitem [xx (1900)]{bb}Woosley, S. E., \& Weaver, T. A. 1994, \apj, 423, 371
\end{thebibliography}
\end{document}